\documentclass[aps,pra,twocolumn,showpacs,preprintnumbers,amsmath,amssymb,superscriptaddress,longbibliography]{revtex4-1}

\usepackage{graphicx}
\usepackage{dcolumn}
\usepackage{bm}

\begin{document}


\title{Generalized framework for applying the Kelly criterion to stock markets}

\author{Tim Byrnes}
\email{tim.byrnes@nyu.edu}
\affiliation{New York University Shanghai, 1555 Century Ave, Pudong, Shanghai 200122, China}
\affiliation{State Key Laboratory of Precision Spectroscopy, School of Physical and Material Sciences, East China Normal University, Shanghai 200062, China}
\affiliation{NYU-ECNU Institute of Physics at NYU Shanghai, 3663 Zhongshan Road North, Shanghai 200062, China}
\affiliation{National Institute of Informatics, 2-1-2 Hitotsubashi, Chiyoda-ku, Tokyo 101-8430, Japan}
\affiliation{Department of Physics, New York University, New York, NY 10003, USA}

\author{Tristan Barnett}
\affiliation{State Key Laboratory of Precision Spectroscopy, School of Physical and Material Sciences, East China Normal University, Shanghai 200062, China}
\affiliation{New York University Shanghai, 1555 Century Ave, Pudong, Shanghai 200122, China}

\date{\today}

\begin{abstract}
We develop a general framework for applying the Kelly criterion to stock markets.  By supplying an arbitrary probability distribution modeling the future price movement of a set of stocks, the Kelly fraction for investing each stock can be calculated by 
inverting a matrix involving only first and second moments.  The framework works for one or a portfolio of stocks and the Kelly fractions can be efficiently calculated.  For a simple model of geometric Brownian motion of a single stock we show that our calculated Kelly fraction agrees with existing results.  We demonstrate that the Kelly fractions can be calculated easily for other types of probabilities such as the Gaussian distribution and correlated multivariate assets. 
\end{abstract}

\maketitle

\section{Introduction}

The Kelly criterion gives an optimal strategy for the long-term growth in risk-taking games where the player has an advantage \citep{kelly1956new}. The strategy is particularly useful in the context of gambling where the future probabilities are explicitly known and has found numerous applications in many contexts. In the context of stock market investments (also called the optimal growth criterion),  the probability of price movements can only be estimated.  Nevertheless there is an extensive literature on applications of the Kelly criterion in this context, for a review see \citep{maclean2011kelly}. In its most basic form, the Kelly strategy states that one should invest a fraction equal to the ratio of the expected return to the winning return \citep{kelly1956new}.  More precisely, for the case of one stock modelled with geometric Brownian motion, one obtains a Kelly fraction of $ f = \mu/\sigma^2 $, where $ \mu $ is the growth rate and $ \sigma $ is the relative volatility. The strategy is known to be optimal in the sense that it minimizes the time to reach a particular wealth, and dominates over any other strategy \citep{breiman2011optimal,finkelstein1981optimal}.  One of the assumptions of the Kelly result is that repeated investments are made over the long term.  This has spurred on investigations of the applicability of the Kelly criterion for finite time horizons \citep{browne2000can}, continuous time variation \citep{browne1996portfolio,thorp2006kelly}, continuous probability distributions \citep{rotando1992kelly}, incorporating risk management \citep{grossman1993optimal}, and comparisons to mean variance approaches \citep{levy1973stochastic,latane1959criteria}.

In a practical scenario, what is particularly of interest is the multivariate case, where more than one stock is invested simultaneously. Numerous studies have been made to utilize Kelly's ideas to calculate the optimal investment fractions in a portfolio of stocks. While all have the basic strategy of maximizing the long-term growth of the assets, the way that the Kelly optimization is performed varies considerably.   Konno and co-workers \citep{konno1991mean} demonstrate that a portfolio optimization model can be reduced to a linear problem making large scale optimizations possible, although is not explicitly based on a Kelly strategy.   Laureti and co-workers \citep{laureti2010analysis} follow a Kelly strategy but do not provide a simple formula to calculation of Kelly fractions.    In the work of MacLean and co-workers \citep{maclean2004capital}, the focus is more towards incorporating risk control to portfolio management than maximizing growth. Furthermore, Nekrasov \citep{nekrasov2014kelly} has derived a formula from which optimal investment fractions can be calculated, but does not take into volatility of the stocks. What would be desirable is a relatively simple procedure to obtain the optimal Kelly fractions given a multivariate probability model of the stock price movements.  To date, we believe that no consensus has been reached for a simple and correct procedure to apply the Kelly strategy for the general case.

In this paper, we provide a new approach to applying the Kelly criterion to stock market investments.  Our approach is to adhere most closely to Kelly's original approach of maximizing the assets over the long-term, given a probabilistic distribution at each investment round. To do this we need to consider the multivariate continuous probability case, which we derive in Sec. \ref{sec:multiple}.  In principle our approach could be applied to an arbitrary probability distribution, not necessary log-normal as is conventionally assumed, and can be correlated.   For the purpose of illustration we solve some portfolio cases with small numbers of stocks in a portfolio, which gives rise to a simple matrix equation which can be solved to obtain the Kelly fractions.  In Sec. \ref{sec:kellysingle} we explicitly evaluate a formula for the Kelly fraction using a log-normal distribution corresponding to geometric Brownian motion, and show that it is consistent with standard results.  We then extend the technique to multiple stocks in Sec. \ref{sec:kellymultiple} and show that a simple procedure involving matrix inversion can be used to obtain the Kelly fractions taking into account volatility.  In Sec. \ref{sec:conc} we summarize our results and main conclusions.

\section{Kelly criterion for multiple outcomes}
\label{sec:multiple}

We first give a short review of the Kelly strategy for multiple probabilistic outcomes, which will also serve to introduce our notation.

\subsection{Single investment case}

Consider a investor with initial assets $ V_0 $ playing a game with $ M $ outcomes with probabilities $ p(i) $, where $ i = 1,\dots,M $. 
For the $ i $th outcome, the return on the invested amount is $ k(i)$, which can be positive or negative.  Thus in the event of the $ i $th outcome, if the investor put all his assets into the game, the total amount of his assets would be $ V_0 ( 1 + k(i) ) $.  The problem is then to calculate what fraction of his assets he should bet in the game, assuming the remaining assets are unchanged.  After $ N $ plays of the game, the total value of his assets will be
\begin{align}
V_N & = V_0 (1 + f k(1))^{n_1} (1 + f k(2))^{n_2} \dots (1 + f k(M))^{n_M} \nonumber \\
& = V_0 \prod_{i=1}^M (1 + f k(i))^{n_i} 
\label{vnmultiple}
\end{align}
where $ n_i $ is the number of outcomes of the $ i $th event, and $ f $ is the investment fraction with $ 0 \le f \le 1 $, assumed constant throughout.  In the standard Kelly approach to finding $ f $ one optimizes the growth
\begin{align}
G = \lim_{N \rightarrow \infty} \frac{1}{N} \ln \frac{V_N}{V_0} .
\end{align}
Taking derivatives one obtains the condition
\begin{align}
\frac{dG}{df} = \lim_{N \rightarrow \infty} \frac{1}{N V_N} \frac{d V_N}{df} = 0 
\label{derivativegrowth}
\end{align}
Substituting (\ref{vnmultiple}) into (\ref{derivativegrowth}) one obtains
\begin{align}
\frac{1}{N} \sum_{i=1}^M \frac{k(i) n_i}{1 + f k(i)} = 0  .
\end{align}
For large $ N $, we expect that $ n_i/N \approx p(i) $ hence we obtain the criterion \citep{Barnett2010,barnett2011much,laureti2010analysis}
\begin{align}
\sum_{i=1}^M \frac{k(i) p(i)}{1 + f k(i)} = 0 .
\label{kellysingle}
\end{align}

\subsection{Multiple investment case}

We can generalize the above to the case of multiple parallel investments. Consider $ L $ investments made in parallel, with a fraction $ f_l $ of the investor's assets allocated to each.  In this case the total fraction invested is then $ 0 \le \sum_{l=1}^L f_l \le 1 $ and the remaining amount is left as cash.  The return of the $i$th outcome for the $ l $th investment is written $ k_l(i) $.   The various outcomes for all the $ L $ investments is then labeled by the outcome $ \bm{i} = (i_1,i_2,\dots,i_L ) $, which occur with a probability $ p(\bm{i}) \equiv p(i_1, i_2, \dots i_L) $.  In the event of the $ \bm{i} $th outcome his assets would be 
\begin{align}
V_0 ( 1 - \sum_{l=1}^L f_l) + V_0 \sum_{l=1}^L f_l (1+ k_l(i_l)) = V_0  ( 1 + \sum_{l=1}^L f_l k_l(i_l)) ,
\end{align}
where the first term is the uninvested fraction and the second gives the returns on each investment. 
In a similar way to (\ref{vnmultiple}) one can then write down the assets of the investor after $ N $ plays of the game
\begin{align}
V_N = V_0 \prod_{i_1, i_2, \dots, i_L } (1 + \sum_l^L f_l k_l(i_l) )^{n_{\bm{i}} } 
\end{align}
where $ n_{\bm{i}} \equiv n_{i_1 i_2 \dots i_L} $ is the number of times the $ \bm{i} $th outcome has occurred. Since there are $ L $ fractions $ f_l $, we have $ L $ different equations (\ref{derivativegrowth}) where the derivative is taken with respect to $ f_l $.  For the $l$th investment fraction we obtain the constraint
\begin{align}
\sum_{i_1, i_2, \dots, i_L} \frac{ k_l (i_l)  p(i_1, i_2, \dots i_L) }{1 + \sum_{l'=1}^L f_{l'} k_{l'} (i_{l'}) } = 0 
\label{kellymultiple}
\end{align}
where we have assumed that $ N $ is sufficiently large enough that $  n_{\bm{i}}/N \approx p(\bm{i}) $.  This is a set of $ L $ equations that must be solved for the $ L $ unknown fractions $ f_l $.

\subsection{Continuous case}

Our aim will be to apply the Kelly criterion formula (\ref{kellysingle}) and (\ref{kellymultiple}) to stock prices.  Since stock prices are essentially continuous, it is more appropriate to use a continuous probability distribution, rather than the discrete case derived above.  We can immediately write this down making the generalization $ i \rightarrow x $, where $ x $ is a continuous random parameter.  We therefore have for the single investment case
\begin{align}
\int dx \frac{k(x) p(x)}{1 + f k(x)} = 0 
\label{singlestockkelly}
\end{align}
where $ p(x) $ is a continuous probability distribution, $ k(x) $ is the return for the outcome $ x $.  For the multiple investment case we have a set of $ L $ equations
\begin{align}
\int d x_1 \dots d x_L \frac{k_l (x_l) p (\bm{x})}{1 + \sum_{l'=1}^L f_{l'} k_{l'} (x_{l'})} = 0
\label{multiplestockkelly}
\end{align}
where $ p (\bm{x}) \equiv p (x_1, \dots , x_L ) $ is the continuous probability distribution for the $ L $ investments, 
$ k_{l}(x) $ is the return on the $ l$th investment.

\section{The Kelly criterion for a single stock}

\label{sec:kellysingle}

\subsection{Probabilistic model for a single stock}

We now apply the general theory of the previous section to stocks. Say the price of a given stock to be currently $ x^{(0)} $. One popular choice of probability distribution is model the price fluctuations by geometric Brownian motion, where the 
new price after some amount of time is
\begin{align}
x^{(0)} \rightarrow x = x^{(0)}  e^\xi
\end{align}
where $ \xi $ is a random number drawn from a Gaussian distribution.  The form of this ensures that the price always remains positive, as is true of real stock prices.  More specifically, the probability distribution then takes a log-normal form
\begin{align}
p(x) = \frac{x^{(0)} }{\sqrt{2 \pi} \hat{\sigma}} \frac{1}{x} \exp \left( - \frac{(\ln x - \ln x^{(0)}  -\frac{\delta x}{x^{(0)} } + \frac{\hat{\sigma}^2}{2 (x^{(0)} )^2} )^2}{2 (\hat{\sigma}/x^{(0)} )^2} \right) ,
\label{lognormalorig}
\end{align}
where $ \hat{\sigma} $ is the standard deviation and $ \delta x $ is the price shift.  The variables $ x, x^{(0)}, \delta x, \hat{\sigma} $ have units of dollars or any other currency of choice (see Appendix). We can define dimensionless ratios of these variables which characterize the distribution
\begin{align}
 \mu & = \frac{\delta x}{x^{(0)} } \nonumber \\
\sigma & =  \frac{\hat{\sigma}}{x^{(0)} }
\end{align}
which is the expected growth in price and dimensionless volatility respectively.  The form of the log-normal distribution is then equivalently written
\begin{align}
p(x) = \frac{1}{\sqrt{2 \pi} \sigma } \frac{1}{x} \exp \left( - \frac{(\ln x - \ln x^{(0)}  -  \mu+ 
\frac{\sigma^2}{2} )^2}{2 \sigma^2} \right) ,
\label{lognormal}
\end{align}
The average and variance of the above distribution takes the values
\begin{align}
\langle x \rangle & = x^{(0)}  e^\mu \nonumber \\
\text{Var} (x) & = \langle x^2 \rangle - \langle x \rangle^2  \nonumber \\
& =  (x^{(0)})^2 e^{2\mu+ \sigma^2 }   -   (x^{(0)})^2  e^{2\mu} .
\end{align}

The investment return takes a simple form in this case
\begin{align}
k(x) = \frac{x- x^{(0)} }{ x^{(0)}}
\label{returnk}
\end{align}
since $ k $ is defined as the fractional return on investment.

\subsection{Evaluation of the Kelly fraction}

We can now examine first for simplicity the single stock case of the model (\ref{lognormal}) and (\ref{returnk}).  While an exact evaluation of the integral (\ref{singlestockkelly}) is only possible using numerical means, for most cases of interest we may assume that the product of the investment fraction and return will be $ f k(x) \ll 1 $.   We can then approximate (\ref{singlestockkelly}) by expanding the denominator as a Taylor series from which we obtain
\begin{align}
& \int dx k(x) p(x) ( 1- f k(x)) \nonumber \\
& = 
e^{\mu} -1 + f(2e^{\mu} - e^{2\mu + \sigma^2} -1) =0.
\label{expandedkelly}
\end{align}
This immediately gives the optimal Kelly fraction
\begin{align}
f = \frac{e^{\mu} -1}{1+ e^{2\mu+ \sigma^2} - 2 e^{\mu}}  .
\label{kellyfractionsingle}
\end{align}
Assuming $ \mu, \sigma \ll 1 $ we can expand the exponential to give
\begin{align}
f \approx \frac{\mu}{\sigma^2} .
\label{conventionalsingle}
\end{align}
This is in agreement with the standard expression for the Kelly fraction for geometric Brownian motion.

\begin{figure}[t]
\centerline{\includegraphics[width=\columnwidth]{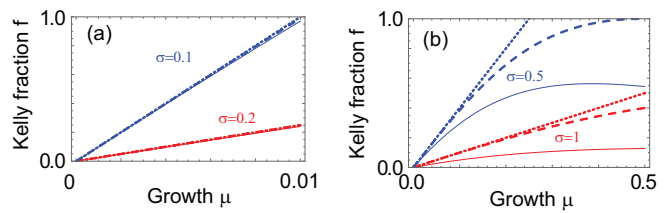}}
\vspace*{8pt}
\caption{Comparison of Kelly fractions versus growth $ \mu $ modeled by geometric Brownian motion using our approach (\ref{kellyfractionsingle}) (solid lines) and the conventional derivation (\ref{conventionalsingle}) (dotted lines) for volatilities as marked. In (a) the volatilities are chosen to be relatively small range, while (b) shows a larger choice volatility.  We also show the Kelly fractions using a Gaussian probability distribution (\ref{gausskelly}) (dashed lines). }
\label{fig1}
\end{figure}

A comparison of our expression (\ref{kellyfractionsingle}) and the standard expression (\ref{conventionalsingle}) is shown in Fig. \ref{fig1}. In Fig. \ref{fig1}(a) we choose relatively small volatilities within the region of approximation of the Taylor expansion performed in (\ref{expandedkelly}).  We see that our Kelly fraction agrees with the standard expression as expected.  This shows that as long as relatively small parameters  $ \mu, \sigma \ll 1 $ are chosen our approach works reliably. 
For a larger choices of $ \sigma $, the Taylor expansion in obtaining (\ref{expandedkelly}) is less valid, and our result starts to differ from the standard result.   However, the curves differ considerably for larger expected growths $ \mu $, generally giving a more conservative Kelly fraction. This is beneficial from the point of view of investing as one would generally like to err on the side of a smaller investment, rather than take on more risk.

To show the general nature of our formalism, we show that it is simple to calculate (\ref{singlestockkelly}) for other choices of probability distribution. Due to the simple form of the return $ k(x) = x/ x^{(0)} - 1 $, one can easily obtain a general formula for the Kelly fraction for an arbitrary probability distribution by evaluating 
(\ref{expandedkelly}):
\begin{align}
f=\frac{\frac{\langle x \rangle}{x^{(0)}} -1}{1+\frac{\langle x^2 \rangle}{(x^{(0)})^2} -2\frac{\langle x \rangle}{x^{(0)}} }
\end{align}
Thus the Kelly fraction can be easily computed from only the first and second moments of the probability distribution.  For example, using a Gaussian distribution (see Appendix) one obtains the Kelly fraction
\begin{align}
f = \frac{\mu}{\mu^2 + \sigma^2} .
\label{gausskelly}
\end{align}
Since $ \frac{\mu}{\mu^2 + \sigma^2} < \frac{\mu}{\sigma^2} $ we again obtain Kelly fractions that are consistently lower than the expression (\ref{conventionalsingle}).  The fractions obtained for the same volatilities are shown in Figure \ref{fig1}.

\section{The Kelly criterion for multiple stocks}

\label{sec:kellymultiple}

\subsection{Kelly fractions for an arbitrary distribution}

We now find the optimal Kelly fractions by substitution of (\ref{multipleprob}) and (\ref{returnkmultiple}) into (\ref{multiplestockkelly}).  To evaluate the integral we can expand the denominator of (\ref{multiplestockkelly}) as a Taylor series to obtain the criterion
\begin{align}
\int d x_1 \dots d x_L k_l (x_l) p (\bm{x}) (1 - \sum_{l'=1}^L f_{l'} k_{l'} (x_{l'})) = 0
\label{multiplestockkelly2}
\end{align}
This is a linear set of equations in $ f_{l'} $ that can be written in matrix form
\begin{align}
{\cal M} \bm{f} = \bm{b}
\end{align}
where we can define the elements as
\begin{align}
{\cal M}_{l l'} & = \int d x_1 \dots d x_L  k_l (x_l) k_{l'} (x_{l'})  p (\bm{x}) \nonumber \\
& = \frac{\langle x_l x_{l'} \rangle}{x^{(0)}_{l} x^{(0)}_{l'}} - \frac{\langle x_l \rangle}{x^{(0)}_{l}} 
- \frac{\langle x_{l'} \rangle}{x^{(0)}_{l'}} + 1 \nonumber \\
\bm{b}_l & = \int d x_1 \dots d x_L k_l (x_l) p (\bm{x}) \nonumber \\
&= \frac{\langle x_l\rangle}{x^{(0)}_{l} }-1  \nonumber \\
\label{matrixdefsgeneral}
\end{align}
and the elements of the vector $ \bm{f} $ are $ f_l $. 
The matrix equation thus only involves the first and second moments of the probability distribution, which are typically easily evaluated.  
To find the optimal Kelly fractions we simply perform a matrix inversion
\begin{align}
\bm{f} = {\cal M}^{-1} \bm{b} .
\label{inversematrix}
\end{align}
Since (\ref{inversematrix}) is a linear equation, this can be computed efficiently.   Eq. (\ref{inversematrix}) is the main result of this paper, and we illustrate this with some simple examples in the following sections.

\subsection{Multiple independent stocks}

To illustrate our main result  (\ref{inversematrix}),  we consider for simplicity the case that the probability distributions for each stock are independent.   In this case the probability distribution takes a form
\begin{align}
p (\bm{x}) = \prod_{l=1}^L p_l (x_l)
\label{multipleprob}
\end{align}
where $ p_l (x_l) $ is the probability distribution for the $ l $th stock. We again take each stock to have a log-normal distribution
\begin{align}
p_l(x) = \frac{1}{\sqrt{2 \pi} \sigma_l } \frac{1}{x} \exp \left( - \frac{(\ln x - \ln x^{(0)}_l - \mu_l + 
\frac{\sigma_l^2}{2} )^2}{2 \sigma_l^2} \right)
\label{lognormalmult}
\end{align}
where we have defined $ \sigma_l = \hat{\sigma}_l / x^{(0)}_l $, $ \mu_l = \delta x_l / x^{(0)}_l $ for the $ l $th stock.  
The returns for each stock is defined as
\begin{align}
k_l(x) = \frac{x-x^{(0)}_l}{x^{(0)}_l} .
\label{returnkmultiple}
\end{align}

Since the probabilities are independent, we can use the fact that the probability distributions are normalized to obtain the simplified expressions for the matrix elements
\begin{align}
{\cal M}_{l l'} & = \left\{
\begin{array}{cc}
A_l & \text{ if $ l = l' $} \\
B_l B_{l'} & \text{ otherwise}\\
\end{array}
\right. \nonumber \\
\bm{b}_l & = B_l 
\label{matrixdefs}
\end{align}
and we have defined the integrals
\begin{align}
B_l & \equiv \int dx k_l(x)  p_l(x) \nonumber \\
A_l & \equiv  \int dx (k_l(x) )^2 p_l(x) .
\end{align}
For the log-normal distribution we can evaluate the integrals to be
\begin{align}
B_l & = e^{\mu_l} - 1 \nonumber\\
A_l & = 1- 2e^{\mu_l} + e^{2\mu_l + \sigma_l^2} 
\end{align}
Alternatively, for the case of independent Gaussian distributions one can evaluate
\begin{align}
B_l  & = \mu_l \nonumber\\
A_l  & = \mu_l^2 + \sigma_l^2  .
\end{align}

\begin{figure}[t]
\centerline{\includegraphics[width=\columnwidth]{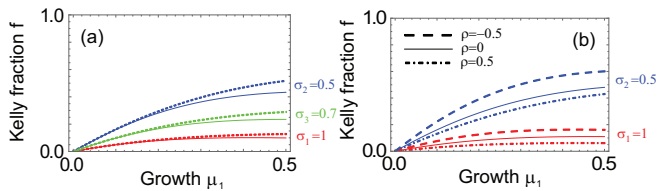}}
\vspace*{8pt}
\caption{(a) Kelly fractions (solid lines) as calculated using (\ref{inversematrix}) using (\ref{matrixdefs}) for $L = 3 $ stocks undergoing independent geometric Brownian motion.  We plot the Kelly fractions as calculated using the single stock formula (\ref{kellyfractionsingle}) for comparison (dotted lines). Parameters used are $ \sigma_1 = 1 $,  $ \sigma_2 = 0.5 $,  
$ \sigma_3 = 0.7 $, $ \mu_2 =  \sigma_2 \mu_1 $, and $ \mu_3 =  \sigma_3 \mu_1 $. (b) Kelly fractions as calculated using (\ref{inversematrix}) using (\ref{matrixdefs}) for $L = 2 $ stocks undergoing correlated geometric Brownian motion. Three values of correlation $ \rho $ are shown as marked. Parameters used are $ \sigma_1 = 1 $,  $ \sigma_2 = 0.5 $,  
$ \mu_2 =  \sigma_2 \mu_1 $.  }
\label{fig2}
\end{figure}

Figure \ref{fig2} shows a sample solution for the Kelly fractions for independent geometric Brownian motion with three stocks. We first note that the calculation of the fractions using  (\ref{inversematrix}) is highly efficient thanks to the exact expressions for the integrals involved in the definition (\ref{matrixdefs}).  Thus while we only show a relatively small example with $ L = 3 $, it is not difficult to perform the computation for a more realistic case of a portfolio involving tens or hundreds of stocks. From Fig.  \ref{fig2} we see that generally the results agree with the single stock formula (\ref{kellyfractionsingle}) for small growths, which in turn agree with the standard expression for the Kelly fraction (\ref{conventionalsingle}).  This is as expected since for small growths only small Kelly fractions would be invested, the effect of having multiple stocks should be not different to the single stock case.  For larger growths, the qualitative effect is to reduce the overall Kelly fraction, which is reasonable from the point of view of reducing the total amount invested. 

 We note that for particular parameter choices, it is possible to obtain Kelly fractions that exceed 1, as is already evident from Fig. \ref{fig1}.  This is also true of the standard solution for the single stock $ f = \mu/\sigma^2 $.  In general (\ref{multiplestockkelly2}) is only valid for small parameters $\mu_l,\sigma_l$ as for the single stock case examined earlier.  For larger values of these parameters, the Kelly fractions must be found by solving (\ref{multiplestockkelly}) directly. However, since in realistic scenarios one typically only considers large $\mu_l,\sigma_l$ for very long-term investments, this should be a reasonable approximation in most cases.

\subsection{Correlated stocks}

More realistically the fluctuations of stocks are correlated. To illustrate our formalism in this case, we show another example for two correlated stocks.  We take the bivariate log-normal distribution \citep{aitchison1957lognormal}
\begin{align}
&p (x_1,x_2) =  \frac{1}{2 \pi \sigma_1 \sigma_2 \sqrt{1-\rho^2}} \frac{1}{x_1 x_2} \nonumber \\
& \times \exp \Big[ - \frac{(\ln x_1 - \ln x^{(0)}_1 - \mu_1 + \frac{\sigma_1^2}{2} )^2}{2 (1 - \rho^2) \sigma_1^2}  \nonumber \\
& + \frac{\rho (\ln x_1 - \ln x^{(0)}_1 - \mu_1 + \frac{\sigma_1^2}{2} )(\ln x_2 - \ln x^{(0)}_2 - \mu_2 + \frac{\sigma_2^2}{2} )}{(1 - \rho^2) \sigma_1 \sigma_2} \nonumber \\
&  - \frac{(\ln x_2 - \ln x^{(0)}_2 - \mu_2 + \frac{\sigma_2^2}{2} )^2}{2 (1 - \rho^2) \sigma_2^2}
 \Big]
\label{bivariateprob}
\end{align}
The average, variance, and covariance of the above distribution takes the values
\begin{align}
\langle x_l \rangle & = x_l^{(0)}  e^{\mu_l} \nonumber \\
\text{Var} (x_l) & = \langle x_l^2 \rangle - \langle x_l \rangle^2  \nonumber \\
& =  (x^{(0)}_l)^2 e^{2\mu_l+ \sigma_l^2 }  -   (x^{(0)}_l)^2  e^{2\mu_l}  \nonumber \\
\text{Cov} (x_1,x_2) & = \langle x_1 x_2\rangle - \langle x_1 \rangle \langle x_2 \rangle  \nonumber \\
&  = x_1^{(0)} x_2^{(0)} e^{\mu_1 + \mu_2} (e^{\rho \sigma_1 \sigma_2} - 1)
\end{align}
for $ l \in \{1,2 \} $.  The returns of the stocks are the same as (\ref{returnkmultiple}).

Using (\ref{matrixdefsgeneral}) we may evaluate the elements of $ \cal M $ and $ \bm{b} $ as 
\begin{align}
{\cal M}_{ll} & = 1 - 2 e^{\mu_l} + e^{2 \mu_l + \sigma_l^2} \nonumber \\
{\cal M}_{12} & = {\cal M}_{21} = e^{\mu_1 + \mu_2 + \rho \sigma_1 \sigma_2 } - e^{\mu_1} - e^{\mu_2}  + 1 \nonumber \\
b_l & = e^{\mu_l} - 1 .
\end{align}

The effect of introducing correlations is illustrated in Fig. \ref{fig2}(b).  We observe that if the stocks prices are positively correlated, the Kelly fractions are reduced overall, and with negative correlation, Kelly fractions are increased overall.  This is the expected result in the context of diversification of assets: negatively correlated stocks give a reduction in risk since the losses of one stock will be offset by the gains of the other.  Positively correlated stocks give effectively more of the same type of stock, hence the Kelly fractions should accordingly be decreased. 

In the limit of two identical and correlated stocks, i.e. $ \mu_1 = \mu_2 = \mu $, $ \sigma_1 = \sigma_2 = \sigma $, and $ \rho = 1 $, we have
\begin{align}
{\cal M} & =( e^{2\mu + \sigma^2} - 2 e^\mu + 1 )\left(
\begin{array}{cc}
1 & 1 \\
1 & 1 
\end{array}
\right)  \nonumber \\
\bm{b} & = (e^\mu - 1) \left(
\begin{array}{c}
1 \\
1
\end{array}
\right)
\end{align}
The matrix $ \cal M $ has eigenvectors $ (1, \pm 1)^T $, with $ (1, - 1)^T $  have a zero eigenvalue.  This makes $ \cal M $  a singular matrix, and is generally not invertible.  However, since $ \bm{b} $ involves only the non-singular eigenvalue, we can perform the inversion (\ref{inversematrix})  and obtain
\begin{align}
\bm{f} = \frac{e^\mu - 1}{2(e^{2\mu + \sigma^2} - 2 e^\mu + 1)}  \left(
\begin{array}{c}
1 \\
1
\end{array}
\right) .
\end{align}
This has exactly half the Kelly fraction of (\ref{kellyfractionsingle}).  This is the expected result as two identical perfectly correlated stocks must give the same Kelly fraction as a single stock, since this is effectively an identical investment.

\section{Conclusions and outlook}

\label{sec:conc}

We have developed a formalism for applying a Kelly strategy to an arbitrary probability distribution for stock market investments. Our formalism is general in that it can be applied to single or multiple stocks in a relatively simple way. Our main result is Eq. (\ref{inversematrix}) where the Kelly fractions can be found by a simple matrix inversion consisting of first and second moments of a probability distribution -- a relatively simple calculation that can be evaluated analytically for typical distributions. For the single stock case with geometric Brownian motion, we reproduce the standard Kelly fraction in the limit of small growth $\mu $ and relative volatility $ \sigma $. Even outside the strict region of validity our calculated results give results that are similar or underestimating the Kelly fraction, which should make it a useful approach in general.

We have shown some prototypical examples to illustrate our formalism for multiple stocks with different parameters, with and without correlation. For multiple independent stocks we observe a smooth crossover between the single stock Kelly fraction at low growths to multiple stock Kelly fractions at larger growths.  For correlated stocks, our results agree with standard investment strategies regarding diversification including negative correlations. These simple tests show that our methods are simple and effective methods for determining the optimal investment fractions, given estimates of key parameters.   An interesting case is the behavior in the regime of large $\mu,\sigma $, corresponding to long-term investments, where the Taylor expansion breaks down. While we generally observed a trend of underestimating the Kelly fraction, a more detailed understanding is needed, particularly for the multiple stock case.

\section*{acknowledgements}
The authors thank Paul R. J. Graham and Saori Katsumata for discussions.  T. Byrnes is supported by the New York University Global Seed Grants for Collaborative Research, National Natural Science Foundation of China (Grant 61571301), the Thousand Talents Program for Distinguished Young Scholars (Grant D1210036A), and the NSFC Research Fund for International Young Scientists (Grant 11650110425), NYU-ECNU Institute of Physics at NYU Shanghai, the Science and Technology Commission of Shanghai Municipality (Grant 17ZR1443600), and the China Science and Technology Exchange Center (NGA-16-001).

\appendix

\section{Limiting case of Log-normal distribution}

The variables in (\ref{lognormal}) are defined such that in the limit that the price fluctuations are small $ | \frac{x-x^{(0)}}{x^{(0)}} | \ll 1 $, we can expand the log-normal distribution to a Gaussian distribution according to the approximation 
\begin{align}
\ln x \approx \ln (x^{(0)} (1+ \frac{x-x^{(0)}}{x^{(0)}} )) \approx \ln x^{(0)} + \frac{x-x^{(0)}}{x^{(0)}}
\end{align}
which then gives
\begin{align}
p (x) \approx \frac{1}{\sqrt{2 \pi} \hat{\sigma}} \exp \left( - \frac{( x - x^{(0)} - \delta x)^2}{2 \hat{\sigma}^2} \right) .
\label{gaussiandist}
\end{align}
The new price will then have a distribution that is centered around $ x^{(0)} + \delta x $ with volatility $ \hat{\sigma} $.

%
%
%


\begin{thebibliography}{18}
\expandafter\ifx\csname natexlab\endcsname\relax\def\natexlab#1{#1}\fi
\expandafter\ifx\csname bibnamefont\endcsname\relax
  \def\bibnamefont#1{#1}\fi
\expandafter\ifx\csname bibfnamefont\endcsname\relax
  \def\bibfnamefont#1{#1}\fi
\expandafter\ifx\csname citenamefont\endcsname\relax
  \def\citenamefont#1{#1}\fi
\expandafter\ifx\csname url\endcsname\relax
  \def\url#1{\texttt{#1}}\fi
\expandafter\ifx\csname urlprefix\endcsname\relax\def\urlprefix{URL }\fi
\providecommand{\bibinfo}[2]{#2}
\providecommand{\eprint}[2][]{\url{#2}}

\bibitem[{\citenamefont{Kelly}(1956)}]{kelly1956new}
\bibinfo{author}{\bibfnamefont{J.~L.} \bibnamefont{Kelly}},
  \bibinfo{journal}{Bell Labs Technical Journal} \textbf{\bibinfo{volume}{35}},
  \bibinfo{pages}{917} (\bibinfo{year}{1956}).

\bibitem[{\citenamefont{MacLean et~al.}(2011)\citenamefont{MacLean, Thorp, and
  Ziemba}}]{maclean2011kelly}
\bibinfo{author}{\bibfnamefont{L.~C.} \bibnamefont{MacLean}},
  \bibinfo{author}{\bibfnamefont{E.~O.} \bibnamefont{Thorp}}, \bibnamefont{and}
  \bibinfo{author}{\bibfnamefont{W.~T.} \bibnamefont{Ziemba}},
  \emph{\bibinfo{title}{The Kelly capital growth investment criterion: Theory
  and practice}}, vol.~\bibinfo{volume}{3} (\bibinfo{publisher}{world
  scientific}, \bibinfo{year}{2011}).

\bibitem[{\citenamefont{Breiman et~al.}(2011)}]{breiman2011optimal}
\bibinfo{author}{\bibfnamefont{L.}~\bibnamefont{Breiman}} \bibnamefont{et~al.},
  in \emph{\bibinfo{booktitle}{The Kelly Capital Growth Investment Criterion:
  Theory and Practice}} (\bibinfo{year}{2011}), pp. \bibinfo{pages}{47--60}.

\bibitem[{\citenamefont{Finkelstein and
  Whitley}(1981)}]{finkelstein1981optimal}
\bibinfo{author}{\bibfnamefont{M.}~\bibnamefont{Finkelstein}} \bibnamefont{and}
  \bibinfo{author}{\bibfnamefont{R.}~\bibnamefont{Whitley}},
  \bibinfo{journal}{Advances in Applied Probability}
  \textbf{\bibinfo{volume}{13}}, \bibinfo{pages}{415} (\bibinfo{year}{1981}).

\bibitem[{\citenamefont{Browne}(2000)}]{browne2000can}
\bibinfo{author}{\bibfnamefont{S.}~\bibnamefont{Browne}},
  \bibinfo{journal}{Finding the edge, mathematical analysis of casino games}
  pp. \bibinfo{pages}{215--231} (\bibinfo{year}{2000}).

\bibitem[{\citenamefont{Browne and Whitt}(1996)}]{browne1996portfolio}
\bibinfo{author}{\bibfnamefont{S.}~\bibnamefont{Browne}} \bibnamefont{and}
  \bibinfo{author}{\bibfnamefont{W.}~\bibnamefont{Whitt}},
  \bibinfo{journal}{Advances in Applied Probability}
  \textbf{\bibinfo{volume}{28}}, \bibinfo{pages}{1145} (\bibinfo{year}{1996}).

\bibitem[{\citenamefont{Thorp}(2006)}]{thorp2006kelly}
\bibinfo{author}{\bibfnamefont{E.~O.} \bibnamefont{Thorp}},
  \bibinfo{journal}{Handbook of asset and liability management}
  \textbf{\bibinfo{volume}{1}}, \bibinfo{pages}{385} (\bibinfo{year}{2006}).

\bibitem[{\citenamefont{Rotando and Thorp}(1992)}]{rotando1992kelly}
\bibinfo{author}{\bibfnamefont{L.~M.} \bibnamefont{Rotando}} \bibnamefont{and}
  \bibinfo{author}{\bibfnamefont{E.~O.} \bibnamefont{Thorp}},
  \bibinfo{journal}{American Mathematical Monthly} pp.
  \bibinfo{pages}{922--931} (\bibinfo{year}{1992}).

\bibitem[{\citenamefont{Grossman and Zhou}(1993)}]{grossman1993optimal}
\bibinfo{author}{\bibfnamefont{S.~J.} \bibnamefont{Grossman}} \bibnamefont{and}
  \bibinfo{author}{\bibfnamefont{Z.}~\bibnamefont{Zhou}},
  \bibinfo{journal}{Mathematical finance} \textbf{\bibinfo{volume}{3}},
  \bibinfo{pages}{241} (\bibinfo{year}{1993}).

\bibitem[{\citenamefont{Levy}(1973)}]{levy1973stochastic}
\bibinfo{author}{\bibfnamefont{H.}~\bibnamefont{Levy}},
  \bibinfo{journal}{International Economic Review} pp.
  \bibinfo{pages}{601--614} (\bibinfo{year}{1973}).

\bibitem[{\citenamefont{Latane}(1959)}]{latane1959criteria}
\bibinfo{author}{\bibfnamefont{H.~A.} \bibnamefont{Latane}},
  \bibinfo{journal}{Journal of Political Economy}
  \textbf{\bibinfo{volume}{67}}, \bibinfo{pages}{144} (\bibinfo{year}{1959}).

\bibitem[{\citenamefont{Konno and Yamazaki}(1991)}]{konno1991mean}
\bibinfo{author}{\bibfnamefont{H.}~\bibnamefont{Konno}} \bibnamefont{and}
  \bibinfo{author}{\bibfnamefont{H.}~\bibnamefont{Yamazaki}},
  \bibinfo{journal}{Management science} \textbf{\bibinfo{volume}{37}},
  \bibinfo{pages}{519} (\bibinfo{year}{1991}).

\bibitem[{\citenamefont{Laureti et~al.}(2010)\citenamefont{Laureti, Medo, and
  Zhang}}]{laureti2010analysis}
\bibinfo{author}{\bibfnamefont{P.}~\bibnamefont{Laureti}},
  \bibinfo{author}{\bibfnamefont{M.}~\bibnamefont{Medo}}, \bibnamefont{and}
  \bibinfo{author}{\bibfnamefont{Y.-C.} \bibnamefont{Zhang}},
  \bibinfo{journal}{Quantitative Finance} \textbf{\bibinfo{volume}{10}},
  \bibinfo{pages}{689} (\bibinfo{year}{2010}).

\bibitem[{\citenamefont{MacLean et~al.}(2004)\citenamefont{MacLean, Sanegre,
  Zhao, and Ziemba}}]{maclean2004capital}
\bibinfo{author}{\bibfnamefont{L.~C.} \bibnamefont{MacLean}},
  \bibinfo{author}{\bibfnamefont{R.}~\bibnamefont{Sanegre}},
  \bibinfo{author}{\bibfnamefont{Y.}~\bibnamefont{Zhao}}, \bibnamefont{and}
  \bibinfo{author}{\bibfnamefont{W.~T.} \bibnamefont{Ziemba}},
  \bibinfo{journal}{Journal of Economic Dynamics and Control}
  \textbf{\bibinfo{volume}{28}}, \bibinfo{pages}{937} (\bibinfo{year}{2004}).

\bibitem[{\citenamefont{Nekrasov}(2014)}]{nekrasov2014kelly}
\bibinfo{author}{\bibfnamefont{V.}~\bibnamefont{Nekrasov}}
  (\bibinfo{year}{2014}).

\bibitem[{\citenamefont{Barnett}(2010)}]{Barnett2010}
\bibinfo{author}{\bibfnamefont{T.}~\bibnamefont{Barnett}},
  \bibinfo{journal}{Law, Probability \& Risk} \textbf{\bibinfo{volume}{9}},
  \bibinfo{pages}{139} (\bibinfo{year}{2010}).

\bibitem[{\citenamefont{Barnett}(2011)}]{barnett2011much}
\bibinfo{author}{\bibfnamefont{T.}~\bibnamefont{Barnett}},
  \bibinfo{journal}{Chance} \textbf{\bibinfo{volume}{24}}, \bibinfo{pages}{10}
  (\bibinfo{year}{2011}).

\bibitem[{\citenamefont{Aitchison and Brown}(1957)}]{aitchison1957lognormal}
\bibinfo{author}{\bibfnamefont{J.}~\bibnamefont{Aitchison}} \bibnamefont{and}
  \bibinfo{author}{\bibfnamefont{J.~A.} \bibnamefont{Brown}}
  (\bibinfo{year}{1957}).

\end{thebibliography}

%
%
\end{document}